# Percolation Phase Transition from Ionic Liquids to Ionic Liquid Crystals

Shen Li[1,2] and Yanting Wang[1,2,]*

[1]*CAS Key Laboratory of Theoretical Physics, Institute of Theoretical Physics, Chinese Academy of Sciences, 55 East Zhongguancun Road, P. O. Box 2735, Beijing, 100190 China*

[2]*School of Physical Sciences, University of Chinese Academy of Sciences, Beijing 100049, China*

Abstract

As typical complex liquids, ionic liquids (ILs) exhibit phases beyond the description of simple liquid theories. In particular, with an intermediate cationic side-chain length, ILs can form the nanoscale segregated liquid (NSL) phase, which will eventually transform into the ionic liquid crystal (ILC) phase when the side chains are adequately long. However, the microscopic mechanism of this transformation is still unclear. In this work, by means of coarse-grained molecular dynamics simulation, we show that, with increasing cationic side-chain length, some local pieces of non-polar domains are gradually formed by side chains aligned in parallel inside the NSL phase, before an abrupt percolation phase transition happens when the system transforms into the ILC phase, manifesting that it is a critical phenomenon. Percolation phase transition is applied to ILs, providing new insights into many recent observations both in experiments and simulations.

---

* Correspondence and requests for materials should be addressed to Y.W.
(email: wangyt@itp.ac.cn).

Significance

A thorough understanding of the transition from the nanoscale segregation liquid (NSL) phase to the ionic liquid crystal (ILC) phase is essential for the wide applications of ionic liquids(ILs), but traditional analyzing methods for phase transition fail to identify its nature, because ILs, as typical complex liquids, do not have the strict spatial symmetry they require. In this work, we manage to identify the change from NSL to ILC is a critical phenomenon with the same universality as the three-dimensional percolation phase transition. This is not only the first time the percolation theory is applied to IL systems, but also demonstrates the importance of the percolation theory in understanding phase behaviors of complex liquids.

Ionic liquids (ILs), also known as room-temperature molten salts, are composed of bulky organic cations and smaller anions. Because of their dual ionic and organic nature (1) originated from the subtle balance among various interactions, such as Coloumbic, van der Waals (VDW), hydrogen bonding, etc., ILs are typical complex liquids exhibiting abundant novel phase behaviors simple liquids do not have, e.g., working under an external force (2, 3), dissolving as solvents (4) and dissolved as solutes (5, 6). However, the phase behaviors, especially phase transitions, of ILs are still poorly understood. In particular, the nanoscale segregated liquid (NSL) phase can be formed when the cationic side-chain length is intermediate (7, 8), and when the cationic side chain is adequately long (9-14), the NSL phase transforms into the ionic liquid crystal (ILC) phase with some unique features beyond traditional liquid crystals (6, 15). Although it has no doubt that a phase transition should exist between the two phases of ILs, the exact phase transition point and its feature is still unidentified. The difficulty comes from the fact that traditional phase transition identification methods usually measure strict geometric symmetries such as translational and/or orientational orders. Percolation phase transition (16, 17), on the other hand, only monitors the degree of connection among particles, which has been intensively studied and exhibits its usefulness in glass transition (18-20) and conductivity (21, 22), yet it has never been applied to IL systems before.

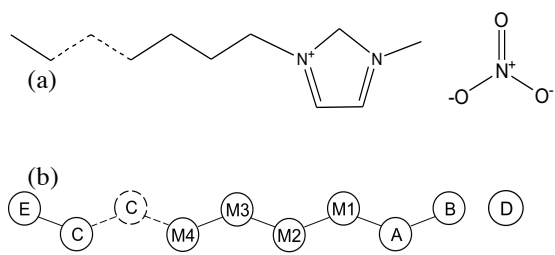

**Figure 1 | Molecular structures of [$C_n$MIM][NO$_3$].** All-atom molecular structure (a) and coarse-grained molecular structure (b) of [$C_n$MIM][NO$_3$]. The cationic imidazolium ring is coarse-grained as CG site A, the single methyl group as CG site B, the tail methyl group as CG site E, CG sites M1, M2, M3 and M4 correspond to the charged methylene groups connected to the imidazolium ring, and CG sites C correspond to all the charge-neutral methylene groups.

In this paper, by performing coarse-grained (CG) molecular dynamics (MD) simulations for [$C_n$MIM][NO$_3$] ($n$ = 12, 14, …, 24, abbreviated as $C_n$ thereafter) ILs at $T$ = 400 K with a simulation size of 4096 ion pairs and an anisotropic barostat allowing the simulated systems to change their sizes in each dimension independently, we manage to identify a sharp percolation phase transition at $n$ = 18. The corresponding physical picture is that, with increasing cationic side-chain length, more local small clusters formed by cationic side chains aligned in parallel appear in the NSL phase, and then go through a sharp percolation phase transition to become the ILC phase when the majority of the cationic side chains are globally aligned in parallel. Therefore, this IL to ILC phase transition is a critical phenomenon, which is difficult or even impossible to be unambiguously identified by common phase transition analysis methods.

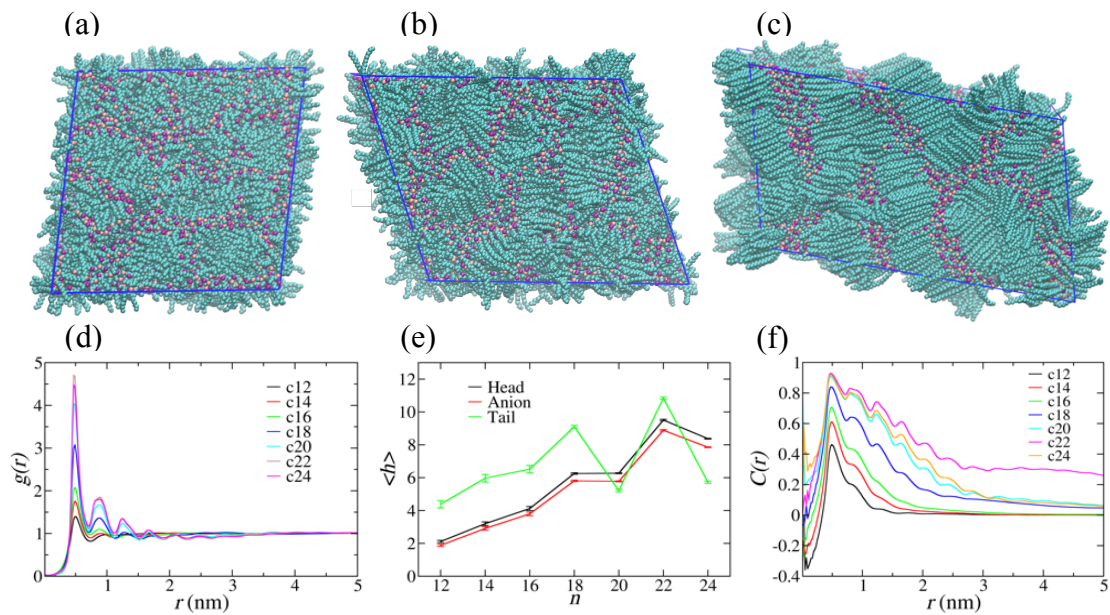

**Figure 2 | Structural properties.** (**a-c**) Randomly chosen snapshots for $C_{12}$, $C_{16}$, and $C_{22}$ systems, respectively. The NSL phase of $C_{12}$ includes a continuous polar network composed of charged groups (anions and cationic head groups) as well as separate nonpolar domains composed of cationic side chains. In $C_{16}$, separate nonpolar domains composed of side-chain groups are locally aligned in parallel without long-range structural order. In $C_{22}$, the side chains are globally aligned in parallel to form the ILC structure. (**d**) RDFs of the side-chain COMs. (**e**) HOPs for anions as well as head and tail groups of cations. (**f**) OCFs for the systems from $C_{12}$ to $C_{24}$. An abrupt transformation can be seen at $C_{18}$ for all these three quantities.

## Results

**Structural properties.** The [$C_n$MIM][$NO_3$] ($n$ = 12, 14, …, 24) IL systems were modeled with the effective force coarse-grained (EF-CG) force field (23, 24), as shown in Figure 1. For each system, 4096 ion pairs were put in a parallelepiped simulation box with the periodic boundary condition (PBC) applied to all three dimensions, and the box size in each dimension changes independently in the *NPT* ensemble. The equilibrium structures at $T$ = 400 K were obtained after appropriate simulated annealing processes (see the Methods section). As shown in Figure 2a, in agreement with the previous studies (7, 8, 25), from the snapshot we can see that $C_{12}$ forms the NSL phase with a continuous polar network composed of charged groups (anions and cationic head groups) and separate nonpolar domains composed of cationic tail groups. From Figure 2b, we can see that the $C_{16}$ system with 4096 ion pairs has only formed some local pieces of cationic side chains aligned in parallel but without long-range structural order, which cannot be regarded as in the ILC phase. By contrast, as can be seen in Figure 2c, the $C_{22}$ system has adequate side chains globally aligned in parallel to form the ILC structure.

To quantify the structural changes, we first calculated the radial distribution functions (RDFs) for the center-of-masses (COMs) of side chains. As shown in Figure 2d, the first peak of the RDF initially increases slowly from $C_{12}$ to $C_{16}$ and then increases quickly from $C_{18}$ to $C_{20}$, before it reaches a quite high value for $C_{22}$. This result indicates that the aggregation degree of cationic side chains increases almost monotonically with side-chain length, and a drastic change happens at $C_{18}$.

The heterogeneity order parameters (HOPs) (26) were then calculated for anions (CG sites D), cationic head groups (CG sites A), and cationic tail groups (CG sites E) to investigate the change of aggregation degree with side-chain length. As can be seen in Figure 2e, the HOP values for anions and cationic head groups overall grows gradually with increasing side-chain length, suggesting that the charged groups become more and more aggregated due to increasing VDW interactions among side chains. In contrast to the higher first RDF peak of $C_{20}$ than $C_{18}$, the HOP values for the charged groups of $C_{18}$ and $C_{20}$ are almost the same, implying that a certain transformation happens at $C_{18}$. The smaller HOP value of $C_{24}$ than $C_{22}$ is very likely attributed to the finite size effect, because the cationic side chains of $C_{24}$ are so long that the current simulation size of 4096 ion pairs is inadequate. Consistent with the RDF results, the HOP of the cationic tail groups increases from $C_{12}$ to $C_{18}$, suggesting that the tail groups become more and more aggregated with increasing side-chain length before the transformation. However, the HOP value for the tail groups of $C_{20}$ is much smaller than both $C_{18}$ and $C_{22}$, while the first COM RDF peak of $C_{20}$ is higher than $C_{18}$. This inconsistency may be explained by the fact that the side chains of $C_{20}$ are more aggregated after the transformation, while the tail groups are in half way of adjusting from separately aggregated nonpolar domains to layered structures, and thus their aggregation degree is less than both structures represented by $C_{18}$ and $C_{22}$, respectively. The decrease of the HOP values for the tail groups of $C_{24}$ is again likely due to the finite-size effect.

The orientation correlation functions (OCFs) (25) were also calculated for the COMs of cationic side chains to quantify their degree of parallelization. As shown in Figure 2f, the first peaks of the OCFs for $C_{12}$ to $C_{24}$ are all located at around 0.5 nm, the distance between two neighboring side chains, so the value of the first OCF peak can quantify the orientation correlation between nearest side chains. The first OCF peak increases with side-chain length from $C_{12}$ to $C_{16}$. The value for $C_{16}$ is around 0.7, a quite strong orientation correlation, indicating that neighboring side chains are well aligned in parallel. However, there are only two peaks formed in the whole range and the OCF value quickly approaches 0 beyond 2 nm, indicating that no long-range orientation correlations exist for side chains. Therefore, from $C_{12}$ to $C_{16}$, with increasing side-chain length, more and more side chains align in parallel locally but no global parallelization is formed. From $C_{18}$ to $C_{24}$, the heights of the first OCF peaks are all around 0.9, demonstrating that neighboring side chains are well aligned in parallel. At the same time, multiple peaks form and the OCF values approach a non-zero value with increasing distance, demonstrating that the systems now have a certain degree of long-range order.

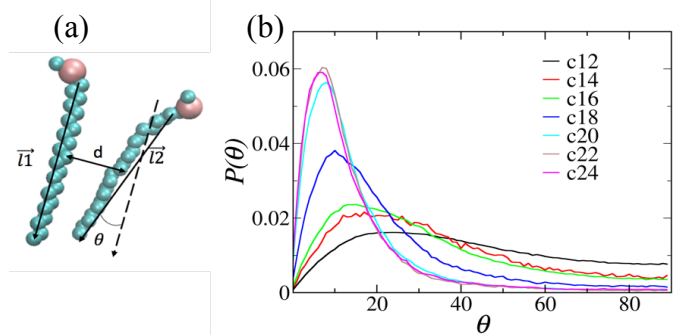

**Figure 3 | Definition of connectivity between cationic side chains and corresponding angle distributions. (a)** The direction of a side chain is defined as the vector pointing from its CG site A (head) to E (tail), *d* is the COM distance between two side chains, and *θ* is the twist angle between two side-chain directions. **(b)** The probability distribution of *θ* is quite flat from $C_{12}$ to $C_{16}$, suggesting a low degree of parallel alignment. In $C_{18}$, the distribution of *θ* is much more tilted to smaller angles than the shorter-chain systems. From $C_{20}$ to $C_{22}$, the distribution of *θ* is mostly populated to angles smaller than 30°, indicating that most side chains are well aligned in parallel.

To have a better understanding of the degree of global parallelization of side chains, the distribution of the angle between two side chains, whose definition is illustrated in Figure 3a, have also been calculated and are plotted in Figure 3b. As shown in Figure 3a, the direction of a side chain is defined as pointing from its head group (CG site A) to its tail group (CG site E).

The angle $\theta$ is the twist angle between two side-chain directions. It can be seen that from $C_{12}$ to $C_{16}$ the distribution of $\theta$ is quite flat, indicating that the orientation correlation between side chains is very weak. The angle distribution of $C_{18}$ has much larger probabilities for smaller angles, and those for $C_{20}$ to $C_{24}$ are mostly populated in the range of 0-30°, indicating that the side chains are better aligned in parallel for those long-chain systems.

Although it is obvious that an abrupt change happens at $C_{18}$ for all the above quantities, none of which can unambiguously identify whether there happens a phase transition at $C_{18}$ or not. In other words, traditional methods involving correlation functions of spatial symmetries cannot characterize the complex phase behaviors of ILs and ILCs. On the other hand, we will show below that a phase transition at $C_{18}$ can be clearly identified by looking at the percolation feature of cationic side chains.

**Percolation phase transition.** A *cluster* is formed by a set of side chains, in which each side chain is "connected" with one or more side chain(s) in the same set. See the Method section for the details of cluster definition. Figures 4a-c show several largest clusters in $C_{12}$, $C_{16}$, and $C_{22}$ systems. Figure 4a shows that only a few small clusters form in $C_{12}$ since the side chains have weak tendencies of aggregation and parallel alignment. As shown in Figure 4b, due to growing abilities of aggregation and parallel alignment, more and larger clusters form locally, but the orientations of the clusters are still random. By contrast, for $C_{22}$ shown in Figure 4c, it is apparent that side chains are in parallel globally and the largest cluster is comparable to the system size, while the second largest cluster is very small.

The normalized sizes of the first and second largest clusters with respect to the side-chain length are plotted in Figure 4d. From $C_{12}$ to $C_{16}$, the sizes of these two clusters are both as small as around 1% and slowly grow with increasing side-chain length. The size of the first largest cluster in $C_{18}$ grows significantly to be around 25%, and the second largest cluster also increases to be around 5%. For the systems from $C_{20}$ to $C_{24}$, the sizes of the largest clusters are all around 70%~80%, comparable to the system size, while the second largest cluster decreases to be around 1%. The slight decrease of the largest cluster size in $C_{24}$ is possibly an artifact caused by the limited simulation size. According to the percolation theory (27, 28), the appearance of a giant cluster suggests that a percolation phase transition happens at $C_{18}$. To verify if the abrupt change at $C_{18}$ is really a percolation phase transition, we have further calculated both the average cluster size (27) (Figure 4e) and the correlation length (27) (Figure 4f) of the side-chain COMs as a function of side-chain length. As shown in Figures 4e and 4f, both the average cluster size and the correlation length reach their maxima of around 500 and 0.42, respectively, at $C_{18}$ with large fluctuations, indicating that a percolation phase transition does happen at $C_{18}$. Consequently, it is clarified that the NSL to ILC phase transition in ILs is a critical phenomenon, i.e., a second-order phase transition.

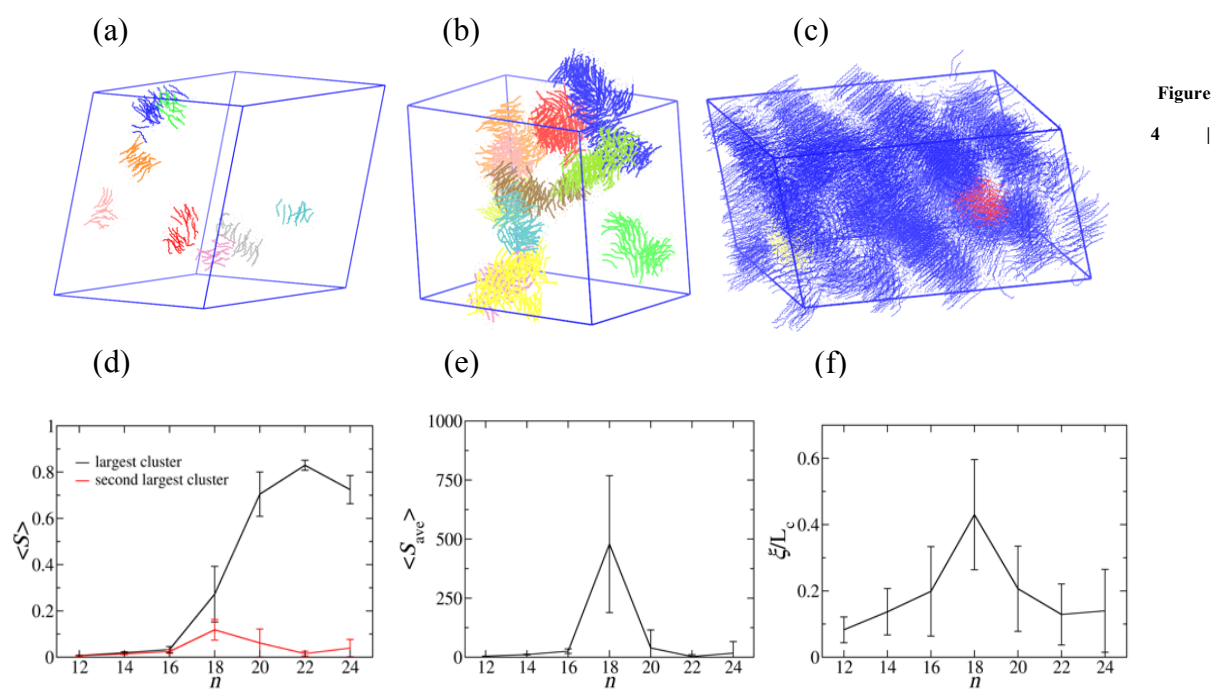

**Percolation phase transition.** (**a-c**) Several largest clusters in $C_{12}$, $C_{16}$ and $C_{22}$ systems, respectively, with the largest cluster colored blue. (**a**) The largest cluster in $C_{12}$ is very small and not well aligned. (**b**) The largest cluster in $C_{16}$ is larger and better aligned in parallel but still local with little orientation correlation between clusters. (**c**) The largest cluster in $C_{22}$ almost fills in the whole simulation box, indicating that the majority of the side chains are globally aligned in parallel and well connected. (**d**) Normalized sizes of the largest and second largest clusters for all systems. (**e**) Average cluster size versus side-chain length. After the percolation phase transition, the largest cluster is not counted in the calculation of the average cluster size. (**f**) Correlation length versus side-chain length. For a finite system, both the average cluster size and the correlation length reach their maxima at the phase transition point. The correlation length is actually directly related to the average cluster size.

## Discussion

In this work, we have performed the CG MD simulations for the $[C_n\text{MIM}][\text{NO}_3]$ ($n$ = 12, 14, …, 24) IL systems with a system size of 4096 ion pairs. To better adapt the asymmetric structural feature of ILC, an anisotropic barostat has been applied to all three dimensions, namely, the simulation box size in each dimension is allowed to change independently in the *NPT* equilibration processes. Our simulation results indicate that, from $C_{12}$ to $C_{16}$, separate small clusters composed of cationic side chains aligned in parallel form locally with the average cluster size gradually increases with the side-chain length, while the whole system structure still retains in the NSL phase. A percolation happens at $C_{18}$ when most of the separate clusters are suddenly unified, corresponding to a large-scale parallel alignment of cationic side chains. The maximization of both the average cluster size and the correlation length as well as their fluctuations manifests that this percolation phenomenon is really a phase transition. The percolation of the cationic side chains in $C_{20}$–$C_{24}$ corresponds to the ILC phase with a global long-range correlation. Although an abrupt change can be seen at $C_{18}$ in RDF, HOP, and OCF plots, only the data analysis method for the percolation phase transition can unambiguously pinpoint the exact phase transition point at $C_{18}$ from the NSL phase to the ILC phase. This critical phenomenon in ILs should have the same universality as the one in the three-dimensional percolation model (16, 17).

Our previous CG MD simulation work (25) has found a sharp transformation around $n = 14$ at $T = 400$ K from the NSL phase to the ILC phase for [$C_n$MIM][NO$_3$] when the cationic side-chain length increases, which qualitatively agrees with experimental observations (9, 13, 14, 29, 30). However, the exact transition point of $n = 14$ was questionable since (1) the simulation size of 512 ion pairs is not large enough to eliminate the finite-size effect, and (2) the isotropic barostat may be inappropriate for ILC simulations. The simulations performed in this work with a much larger size of 4096 ion pairs and an anisotropic barostat clearly shows that the structural change around $n = 14$ becomes mild, and a sharp percolation phase transition appears at $n = 18$.

Some experiments (31, 32) reported similar observations, but the exact transition point may vary according to different conditions, such as temperature, type of ions, etc. The purpose of this work is not pinpointing the exact transition point, rather it provides a qualitative and general theoretical framework for understanding the phase behavior of ILs changing from the NSL phase to the ILC phase with increasing cationic side-chain length, which may be very important for the applications of ILs and ILCs in dye-sensitized solar cells (33), electrofluorescence switches (34), electrolytes for Li-ion batteries (35) and electrochemical sensors (36).

With complex molecular structures and interactions, complex liquids frequently exhibit complex phase behaviors beyond the description of simple liquid theories. Therefore, the identification methods for the phase transition of simple liquids, such as the translation and orientation correlation functions, may not be applicable to determining the phase transitions of complex liquids. In this work, we have demonstrated that, as typical complex liquids, the phase transition of IL systems from the NSL phase to the ILC phase cannot be unambiguously identified by the correlation functions requiring high spatial symmetry due to the complexity of the spatial features of the NSL and ILC phases. By contrast, the data analysis method for percolation easily and clearly identifies the phase transition point, because the percolation phenomenon merely requires a minimal spatial symmetry of connectivity. As the first time that percolation phase transition is applied to IL systems, our results demonstrate the necessity of developing novel liquid theories for complex liquids as well as the importance of percolation theory in describing phase behaviors of complex liquids.

## Methods

**CG MD simulation.** The $[C_n\text{MIM}][NO_3]$ ILs, $n$ = 12, 14, …, 24, are modeled by the EF-CG force field (23, 24). All the simulated systems contain 4096 ion pairs in a parallelepiped box with the PBC applied to all three dimensions. A cutoff distance of 1.4 nm was applied to the VDW and the real part of the electrostatic interactions, and the particle-mesh Ewald method (37) was employed to handle the long-range electrostatic interactions. The temperature was kept constant by using the Nosé-Hoover thermostat (38) with a time constant of 0.5 ps. For the *NPT* simulated annealing processes from 1200 K to 800 K, the isotropic Parrinello-Rahman barostat (39) with a time constant of 2 ps was employed to keep the system pressure constant. For all other *NPT* cases, the anisotropic Parrinello-Rahman barostat with $P$ = 1 atm was applied to the three dimensions independently. All the simulations were performed by using the GROMACS software package (40) with a time step of 4 fs.

For each IL system, a large random initial configuration was first equilibrated by a 1-ns *NPT* MD simulation at *P* = 10 atm and *T* = 100 K to reduce molecular distances, followed by an *NPT* simulated annealing process cooling from 1200 K down to 800 K with a temperature interval of 100 K and 2-ns simulated time at each temperature. For convenience, an isotropic barostat was applied for this annealing process. Another subsequent *NPT* simulated annealing process was performed from 800 K down to 400 K with a temperature interval of 100 K and 8-ns simulated time at each temperature, during which an anisotropic barostat was employed to adjust the size in each dimension independently. The systems finally went through an equilibrium *NPT* (anisotropic barostat) run at *T* = 400 K for 8 ns with 2000 configurations evenly sampled from the simulation trajectory.

**Heterogeneity order parameter.** The HOP(26) continuously quantifying the spatial heterogeneity is defined as

$$h = \frac{1}{N} \sum_{i,j=1}^{N} \exp(-r_{ij}^2 / 2\sigma^2) \quad (1)$$

where $r_{ij}$ is the distance between site *i* and site *j* corrected by the PBC. The normalized distance parameter $\sigma = (V/N)^{1/3}$ with *V* being the system volume and *N* being the total number of sites.

**Orientation correlation function.** The OCF (25) for side chains were calculated to quantify the degree of parallel alignment of side chains. It is defined as the ensemble-averaged orientation correlation between two side chains as a function of their COM distance:

$$C(r) = \left\langle \left[ 3(\hat{u}(\vec{r}_i) \cdot \hat{u}(\vec{r}_j))^2 - 1 \right] \cdot \delta(\vec{r} - \vec{r}_i + \vec{r}_j)/2 \right\rangle \quad (2)$$

where $\hat{u}(\vec{r}_i)$ is the unit vector of cation *i* located at $\vec{r}_i$ pointing from CG site M1 to CG site E.

**Twist angle distribution.** To quantify the degree of global parallelization, the ensemble-averaged twist angle distribution between two side chains were calculated as

$$p(\theta) = \frac{\sum_{i,j} \arccos(|\hat{u}(\vec{r}_i) \cdot \hat{u}(\vec{r}_j)|)}{N(N-1)} \cdot \delta(\theta) \qquad (3)$$

The twist angle distribution is close to uniform if all side chains are randomly oriented or locally aligned in parallel but globally randomly oriented. Only when side chains are globally aligned in parallel can the angle distribution be mostly populated in small angles.

**Definition of cluster.** A cluster is defined as a set of "connected" side chains. The COM distance and the twist angle between two side chains are combined together to identify whether they are connected. The COM RDFs in Figure 2d indicate that the first shell is within 0.72 nm, and the twist angle distribution in Figure 3b shows that the angle is mostly within 30° for high parallelization cases. Therefore, two side chains are considered "connected" if the COM distance is less than 0.72 nm and at the same time the twist angle between the two side chains is less than 30°. In a cluster, each side chain is connected to at least one side chain in the same cluster.

**Average cluster size and correlation length.** The probability that a site belongs to cluster $i$ of size $S_i$ is $S_i/N$, so the probability that a randomly chosen cluster has the size of $S_i$ is $S_i^2/N$. Therefore, the average cluster size $S_{ave}$ of a given system can be calculated by (27)

$$S_{ave} = \frac{1}{N} \sum_{i=1}^{N_c} S_i^2 \qquad (4)$$

where $N_c$ is the total number of clusters. Note that theoretically only sites belong to finite clusters should contribute to the calculation since an infinite cluster would make the average cluster size diverge, so numerically the largest cluster after percolation phase transition is not counted in the calculation of the average cluster size.

The correlation length $\xi$ is defined as the ensemble average of the distance between two sites in the same finite cluster (27)

$$\xi^2 = \frac{2 \sum_i R_i^2 S_i^2}{\sum_i S_i^2} \qquad (5)$$

where the radius of gyration $R_i$ of cluster $i$ is defined as $R_i^2 = \frac{1}{2}\sum_{j,k}(r_j - r_k)^2 / S_i^2$. The correlation length $\xi$ is indeed directly related to the average cluster size $S_{ave}$, because a longer correlation length always corresponds to larger clusters given that only pairs in the same cluster contribute to the correlation length.

## Acknowledgments

The authors thank Prof. Haijun Zhou for his critical reading of this paper. This work was supported by the National Natural Science Foundation of China (Nos. 11774357 and 11747601) and the CAS Biophysics Interdisciplinary Innovation Team Project (No. 2060299). The authors thank Tianhe-2 supercomputer and the HPC Cluster of ITP-CAS for allocations of computer time.

**Author contributions**

S.L. and Y.W. conceived the research, S.L. performed simulations and analysis, S.L. and Y.W. discussed and wrote the manuscript.

**Competing financial interests:** The authors declare no competing financial interests.

**Reprints and permission** information is available online at http://npg.nature.com/reprintsandpermissions

**How to cite this article:** ***